# The Digital Foundation Platform - A Multi-layered SOA Architecture for Intelligent Connected Vehicle Operating System

David Yu, VP of Engineering and Chief Architect, AICC Inc.
Andy Xiao*, Ph.D., Software Architect, AICC Inc.

## Abstract

Legacy AD/ADAS development from OEMs centers around developing functions on ECUs using services provided by AUTOSAR Classic Platform (CP) to meet automotive-grade and mass-production requirements. The AUTOSAR CP couples hardware and software components statically and encounters challenges to provide sufficient capacities for the processing of high-level intelligent driving functions, whereas the new platform, AUTOSAR Adaptive Platform (AP) is designed to support dynamically communication and provide richer services and function abstractions for those resource-intensive (memory, CPU) applications.

Yet for both platforms, application development and the supporting system software are still closely coupled together, and this makes application development and the enhancement less scalable and flexible, resulting in longer development cycles and slower time-to-market.

This paper presents a multi-layered, service-oriented intelligent driving operating system foundation (we named it as Digital Foundation Platform) that provides abstractions for easier adoption of heterogeneous computing hardware. It features a multi-layer SOA software architecture with each layer providing adaptive service API at north-bound for application developers.

The proposed Digital Foundation Platform (DFP) has significant advantages of decoupling hardware, operating system core, middle-ware, functional software and application software development. It provides SOA at multiple layers and enables application developers from OEMs, to customize and develop new applications or enhance existing applications with new features, either in autonomous domain or intelligent cockpit domain, with great agility, and less code through re-usability, and thus reduce the time-to-market.

## 1. Introduction

The automotive industry has been undergoing major disruptive digital transformations. These major transformations are characterized by ACES - Autonomous Driving (Intelligence), Connectivity (Interconnection), Electrification and Shared-Mobility, as shown in Figure 1. There are multiple catalysts that drive and enable these trends.

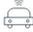

Figure 1. ACES trends in automotive industry, source: McKinsey analysis.

First, collective awareness of environmental and climate concerns. The automotive industry contributes 15 -25 percent of polluting emissions [1], which affect environment pollution and climate change. To address the concerns, traditional fossil-fuel-powered vehicles are increasingly being replaced by electrical vehicles. The electrification of vehicles at the same time accelerates the development of new electrical packages and new power-train applications including complex battery-management software.

Second, shared economies. The biggest upside of shared mobility is financial savings. People no longer need to own cars, and can be served by on-demand vehicles, e.g., ride sharing [2]. On the other hand, shared-mobility can accelerate the introduction of autonomous vehicles due to the cost-efficiency.

Third, user experience. There is an increasing demand for smart driving experience and staying connected. Vehicle interconnection provides benefits of infotainment, traffic safety, on-board diagnostics, etc., and makes life better in a variety of ways. On the other hand Interconnection boosters innovations in communication, e.g., V2V, V2X, and reinforces and accelerates shared-mobility and autonomous driving [3].

Fourth, the huge growth in cloud technology, big data analysis and artificial intelligence (AI). Cloud computing and big data analysis technologies facilitate advanced autonomous driving technologies, for example, more accurate perception and decision making through continuous driving data collection, analysis and fine tuning of AI models to improve accuracy, reduce false positives and address the so called "long tail" problems in AD. The emerging AI technologies including machine learning (ML), deep learning (DL), neural

*Xinhua Xiao, Email: andyxiao@aicc-corp.com

network (NN), and computer vision are moving forward to allow higher levels of AD [4].

With the four digital transformation trends-ACES, the complexity of software built into vehicles is growing exponentially. Add to this the massive explosion of sensor data, and the need to manage, and interpret all this data using AI techniques in real time and using communication technologies to transfer the data.

In addition, an indispensable part of a prospective AD development is communication over cars and infrastructure (connected vehicles) [5-6]. The Connected vehicles (CV) use vehicle-to-vehicle, vehicle-to-infrastructure, and infrastructure-to-vehicle communication to exchange information between vehicles, the roadside, bicyclists and pedestrians. The connectedness of AD vehicles has been investigated in signal control policies and helps significant reductions in delay in traffic flow [6]. However, the connectivity opens up new possibilities for cyber-attacks, including in-vehicle attacks and V2X communication attacks (e.g., data theft) [5, 7], especially with the development of 4G LTE and 5G communication technologies. Many efforts are made to improve the resilience to cyber-attacks, e.g., software vulnerability detection [7] and software testing techniques such as symbolic execution [8] and mutation testing [9].

More and more Information and Communications Technologies (ICTs) are joining the automobile industry revolutions, and ICTs have become the key enabler for these digital transformations, especially in fields of autonomous driving and intelligent connections.

As more and more features and functions are being added to the vehicle, the traditional legacy ECU based E/E vehicle architecture (each ECU is responsible for a single function) will strain under pressures due to space limit. There is a growing demand for consolidation of ECUs into a centralized computer architecture, which in turn enables the so-called Software Defined Vehicle (SDV) [10], allowing new functions through software update over-the-air (OTA).

In this paper, we will mainly focus on the software infrastructure working with Digital Foundation Platform (DFP), a multi-layered Service-Oriented Architecture (SOA) handling software management, critical computing task allocation, data flow handling in the vehicle, and communication with the cloud.

The remaining of the paper is organized as follows: Section 2 describes the key development in modern intelligent connected vehicles (ICVs). Section 3 details the technologies of E/E Architecture, Service Oriented Architecture, and Software Defined Vehicles. Section 6 explains the legacy development problems, and the DFP (Digital Found Platform) is proposed and illustrated in Section 7. Advantages of the DFP are detailed in Section 8. Section 9 gives the conclusion. Future work and existing problems to solve are given in Section 10.

## 2. Key Development in Modern Intelligent Connected Vehicles (ICVs)

### 2.1 E/E Architecture

With the ACES transformations, there is overall trend for the transition of vehicle E/E architecture from highly distributed ECU systems towards centralization. Traditional OEMs' E/E architectures are in "distributed" or "modular" fashion as shown in the bottom part of Figure 2. There might be up to 150 electronic control units (ECUs) in a vehicle, and each ECU is a dedicated chip that combines a processor, memory and software and is designed for controlling a specific function, handling features from infotainment to advanced safety, comfort and connectivity, the so-called "one function - one box". ECUs are interconnected through cables, and there is limited communication between ECUs. As more functions and features are added, especially the autonomous driving and connected features for ICVs under the ACES trends, the traditional architecture will go under challenges, and no longer meet the need due to space, computation and communication limit. To solve this, the E/E architectures needs to be upgraded or redesigned. Take Bosch as example, the evolution of the E/E architecture road-map consists of six phases,

1) Modularization -One ECU corresponds to one function,
2) Integration -ECU begins to integrate multiple functions,
3) Centralization -ECUs consolidates into domain control units (DCUs),
4) Domain fusion -DCUs with similar functions are further integrated,
5) On-board computer -All DCUs are fused into one central vehicle computer,
6) Vehicle-cloud computing -computing is transferred into the cloud server.

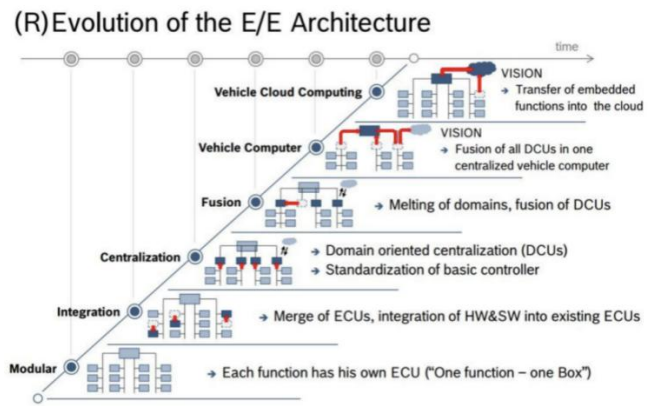

Figure 2. Example of E/E Architecture Evolution. Source: Bosch

### 2.2 Service Oriented Architecture (SOA)

E/E vehicle architectures will become more centralized as it enhances the capabilities of update and upgrade the functions and features for ICVs. Enabled by this transition, Service Oriented Architecture (SOA) has been introduced, allowing dynamic communication and simplified access to information instead of static dependencies mapped on ECUs [11].

SOA is a widely used software development methodology in ICT industry for a long time. It refers to the way to make software components modular and provide common and abstract service interfaces so that these components can be reused, either through Software Development Toolkit (SDK) or software libraries. The interfaces are designed using common standards so that the new applications can be developed atop them in an efficient and agile way, leveraging existing investments in software development.

Following the development rules of the ICT, automotive software development introduces the SOA middleware and virtualization technologies to achieve software modularization and substantial abstraction levels to services and applications. For a typical example of the Service Oriented Architecture(SOA) in ICVs, as shown in Figure 3, servicesSvc1 -Svc4 are provided through SOA Middleware running on top of Automotive Ethernet & TCP/IP [12], and the consumer application on the right part of the Figure can consume services either Srv2 or Srv4 without knowing of the location.

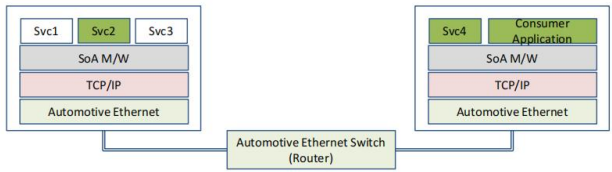

Figure 3.   SOA example in Automotive, source: [12]

## 2.3 Software Defined Vehicles (SDVs)

Software Defined Vehicle (SDV) refers to the vehicle whose functions are primarily enabled through software and can be continuously upgraded. It means that the value of software exceeds the hardware and becomes a bigger part of automobiles [13]. To realize SDVs, the vehicle E/E architecture is moving toward a centralized architecture from consolidating ECUs and functions. This centralized architecture allows better software management and computing efficiency. In addition, it makes possible for feature or function updates Over-the-air (OTA) to distribute to the vehicle the new software code. To enable SDVs, a SOA architecture is adopted to host applications and services to support data transfer and communication, and new functions can be easily updated and added to vehicles.

## 3. E/E Architecture

Distributed ECUs communicate through different bus systems [11, 14-16]. The vehicle bus systems include CAN, Local Interconnect Network (LIN), FlexRay, Media Oriented Systems Transport (MOST) and Automotive Ethernet. CAN is mainly used for signal-oriented communication, and Ethernet is adopted for service-oriented paradigm.

### 1) Runtime Platforms

The software architecture of ECUs is based on OS and runtime stacks [15]. Two typical platforms are AUTOSAR Classic and AUTOSAR Adaptive. The Classic Platform is used for high real-time requirements, while the Adaptive Platform is suitable for greater computing power and is used for SOA paradigm. Besides AUTOSAR, QNX and GENIVI are platforms for infotainment domain. Table 1 sums up the major differences.

Table 1. Comparison of AUTOSAR Classic and Adaptive platform, source: [17].

| Property | Classic Platform | Adaptive Platform |
|---|---|---|
| Operating system | based on OSEK | based on POSIX |
| Code execution | from ROM | loaded into RAM |
| Application address space separation | no | yes |
| Scheduling | fixed | dynamic (preemptive multitasking, varying number of tasks) |
| Compiling | ECU compiled as whole | Applications are installed in POSIX processes |
| Configuration | Compiled system configuration | Loaded at runtime from file |
| Supported protocols / paradigms | signal-oriented | signal-oriented, SOA (DDS, SOME/IP) |

From Table 1, AUTOSAR Adaptive is more dynamic, and scheduling is dynamic for applications.

### 2) Future Architecture

The trend of ACES leads to centralization and stronger computing platforms [11]. E/E architecture might run on both CLASSIC Platform and AUTOSAR Adaptive. Figure 4 shows an exemplary E/E architecture of a future vehicle. The central computing cluster is used for main computation and runs AUTOSAR Adaptive. The connectivity control and I/O cluster are also based on AUTOSAR Adaptive. The chassis and ADAS functions are safety critical and run AUTOSAR Classic. For functions with low data rates like Body and Infotainment, the cheaper CAN/LIN is used instead of the Ethernet.

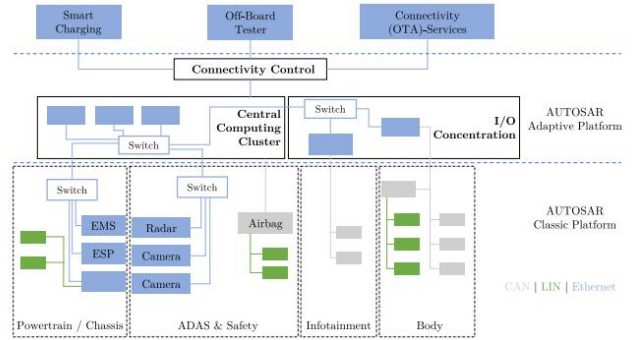

Figure 4. Exemplary hybrid E/E architecture based on AUTOSAR Classic & Adaptive. ECU colors indicate the respective bus systems they are attached to, source: [17].

## 4. Service Oriented Architecture (SOA)

One of the important aspects of SOA is cross platform support. With abstractions and standardized services interfaces SOA [18-19], it is easy to decouple the application development and underlying system or platform support. The following presents how to absorb change and support cross platform, and modularized SOA and generalized computing platform.

### 4.1 Cross Platforms

There used to be a wide varieties and number of ECUs in the vehicle, but for different applications, they often run on different OSs and devices. For example, the software for object recognition in front view runs on the device R-car/V3H with QNX OS shown in the left section of Figure 5, while the one for object recognition in surround view runs on the device R-car/H3 with Linux OS [20]. The two functions are similar, and to remove the variant developments and maximize the reuse of software, Renesas introduces abstraction layers that absorb the differences between HW, OS and middleware. There are three such abstraction layers, including Hardware abstraction layer (HAL, which makes common driver and common Micro controller Abstraction Layer -MCAL for all devices), OS abstraction layer (OSAL, which allow the same middleware on all types of OS), and the standardized API.  These abstraction layers are the essential building blocks to create a common/cross platform which is independent from middleware, OS and device variants. The new software platform enables cross platform, and significantly improves software re-usability.

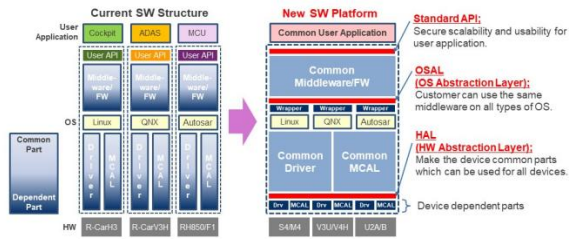

Figure 5.   Cross-Platform, source: Renesas [20].

## 4.2 Modularized SOA and Generalized Computing Platform

Figure 6 shows a SOA based on a layered classification of ECUs. Components serving certain vehicle functions will be encapsulated in layers. Three layers of services are designed, a basic layer for services such as processing sensor data, an extended middleware layer for data exchange and data fusion, and an application layer using all this data for applications such as brake or side-view assistant. Encapsulation and layering reduce fragmentation, complexity and facilitate the using of generalized computing platform.

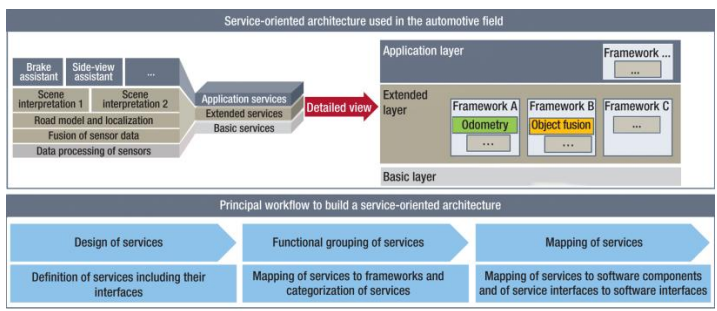

Figure 6.   SOA example in the automotive, source: BMW [15].

## 5. Software Defined Vehicle (SDV)

Software-driven transformation is truly one of the key technologies that is reshaping the industry. Software is the key to developing vehicles to meet the ACSE trends. It is estimated that for level 5 autonomous driving, the number of lines of codes can exceed one billion [7]. The key for SDV is that the vehicle is not deeply tied by the hardware; rather, software mainly defines the features, functionalities, and in-vehicle user experience [21]. To achieve this, a diverse array of ICT technologies is needed including hardware abstraction, middleware, virtualization, and containerization. Figure 7 shows an example of SDV technology stack. On the bottom, we can see the E/E architecture is centralized, with powerful High-Performance Computer (HPC) as the computing units. On top of it is the vehicle platform which consists of OS, hypervisor and middleware, which serves the key part to separate the dynamic software stack including services and applications from the vehicles.

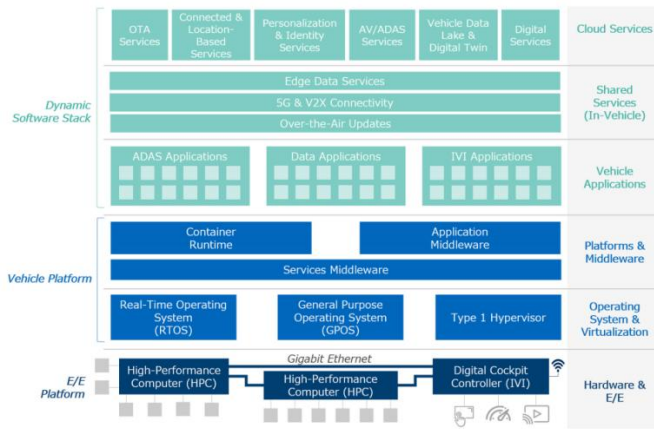

Figure 7. The Software-Defined Vehicle Technology Stack, source: [21].

## 6. Legacy Development Problems

The legacy distributed ECUs based E/E architectures common today rely on signal-based communication and have proven reliable and secure for OEMs. However, with the ACES trends, the legacy development methodologies will strain under the pressures and challenges as following.

### 6.1 Function Complexity

The software each ECU resides on varies greatly, and the computing capability of the ECU is limited. OEMs often acquires ECUs as black boxes with limited insight into the software. As more functions such as autonomous driving and connectivity features are added into the vehicle, more ECUs are needed and it requires huge amounts of integrating these functions, and also the complexity of physical wiring and network harness increases. Besides, to track all legacy electronic updates and software in each model and manage vehicle variant become difficult for OEMs.

### 6.2 Testing Challenges

To bring all ECUs together and testing combination of the ECUs and the communication between them need significant investment and effort.  It is estimated by Deloitte that system integration, testing, verification and validation accounts more than 40% vehicle development cost [13]. Vehicle models differ in configurations and features, and before rolling off the production it needs thorough testing. There might be all kinds of electronic combinations, and millions or billions of test set-ups might be needed. To reduce the set-ups and build common boards/ digital wheels during vehicle development are necessary.

### 6.3 Software Upgrade

Since in legacy systems, software and underlying hardware are tightly coupled together, any software changes, be it bug fixing, feature enhancement or module upgrade, will be difficult, resources consuming and error prone. This is one of the most important reasons that traditional OEMs are obligated and motivated to make strategy changes to migrate to software defined and service-oriented system and software architecture.

# 7. The Digital Foundation Platform

To address the problems of the legacy hardware and software development problems in OEMs and Tier1 manufactures, we propose a multi-layered, highly flexible, adaptive software architecture for ICVs.The overall architecture is shown in Figure 8.

It is a multi-layer architecture that consists of the following layers including hardware platform layer, system software, middleware, function software and application software.

## 7.1 Hardware Platform Layer

For L3 and higher-level autonomous driving vehicles, the basic platform for on-board AI computing needs to support different types and numbers of sensors with high safety and performance requirements. The single chip solution can no longer satisfy many different interfaces, and not mention support complex deep learning and other AI algorithm computing requirements, instead, heterogeneous chip and hardware solutions are to be used.

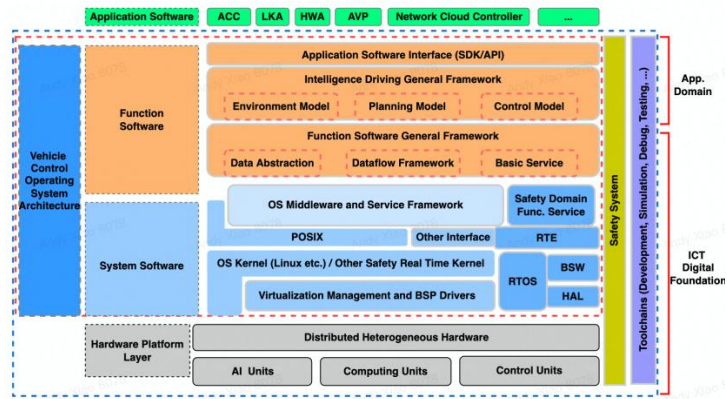

Figure 8. Proposed Digital Foundation Platform -A multi-layered SOA Architecture for Intelligent Connected Vehicles.

Hardware heterogeneity can refer to single-board card integration, multiple chips architecture, such as integrated MCU, FPGA, CPU etc.; it can also be embodied in a powerful SOC integrates multiple architecture units at the same time, such as the NVIDIA Xavier which has two heterogeneous units - GPU and CPU.

The basic heterogeneous hardware composition usually includes AI units, computing units and control units. AI units have parallel computing architecture and use multi-core CPU or GPU, FPGA, ASIC, etc. They are used to conduct the most computing intensive tasks such as raw camera image data processing or early multi-sensor data fusion etc., thus require high performance. Specially built hardware acceleration engines are used to accelerate the AI tasks.

For example, the Mobile Data Center (MDC) from Huawei uses what it calls "mini" ASIC on its model 300 to accelerate the image processing data received from camera sensors. Computing unit uses automotive grade multicore CPU chip sets with higher single chip computing capacity. Most of the fundamental autonomous driving algorithms including perception, localization, fusion, planning algorithms. There are other non-autonomous driving services are also running on the computing unit including data management, vehicle cloud collaboration etc. Software running on the computing unit need to meet different levels of functional safety requirements.

Typical autonomous driving computing units include Mobileye's EQ series, Nvidia Jetson AGX Xavier or Orin, etc.

Control unit refers to the traditional vehicle control on MCU. It uses AUTOSAR CP to provide abstractions and services for software development on ECUs. Control unit connects to ECU via vehicle control interfaces to conduct longitudinal and lateral control of the vehicle as well as other vehicle dynamics. Software running control unit must meet ASIL-D functional safety requirements.

Hardware on an autonomous system come with varieties of forms, sensors, actuators, and vehicle electric and dynamics. Here we abstracted the hardware and adapt our DFP to different hardware platforms, including MDC, Nvidia Xavier, Ti Chips, etc.

## 7.2 System Software

System software runs atop the heterogeneous hardware, through Hardware Abstraction Layer (HAL), it hides the detailed differences of these hardware elements. System software is one of the core components of the autonomous driving foundation platform. System software runs on both ADAS/AD computing domain as well as safety domain.

### 7.2.1 Virtualization

In the system software layer, virtualization management is used as hypervisor or virtual machine monitor into the system and runs virtual machines (VMs). For high levels of performance, reliability and security capabilities in platform virtualization, Type Zero hypervisor [22] is used as shown in Figure 9. The Type Zero hypervisor is a bare-metal architecture built with the minimum software components required to fully virtualize guest operating systems and control information flow between guest operating systems [23]. The Type 0 architecture removes the need for an embedded host OS to support virtualization, allowing the hypervisor to run in an "Un-Hosted" environment [22-23]. The hypervisor handles and controls accesses to the virtualized resources (the hardware). These virtual machines host OS kernels including Android, QNX, Linux, real-time OS AUTOSAR. Strong isolation between individual VMs is enforced, and ensures safety and security. Software containers are created and the vehicle domains are isolated, not affecting each other. The OS kernels implement the standard POSIX API. With those standardized POSIX kernels, the vulnerabilities are introduced. To ensure the safety of the platform and meet the functional safety requirements, MCUs with Safe RTOS should be integrated.

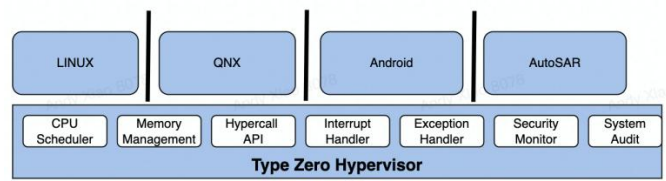

Figure 9. Type Zero Hypervisor, a bare-metal architecture run in an unhosted environment.

### 7.2.2 Safe RTOS

A Real Time Operating System (RTOS) provides rigorous resource management and scheduling to meet the demands of functional safety. RTOS is provided for MCU to meet functional safety requirements. A set of safety requirements are implemented, and safety domain function services are provided. It supports AUTOSAR, advanced vehicle dynamic control and fail-safe automated driving functions, and implements arbitration control strategies.

## 7.3 Middleware

Middleware lays on top of the system software. It controls the data flow and performs type checking. Middleware usually consists of communication sublayer and services. For automotive SOA, important middleware communications usually use DDS mechanism and use SOME/IP [24] and other communication protocols. SOME/IP specifies [17] "Yet another RPC-Mechanism" in automotive by AUTOSAR. It simultaneously creates abstraction between the application and the network at runtime instead of at the system design time. DDS is a data centric middleware based on the publish-subscribe pattern to control the flow of data between different nodes. It uses real time publish subscribe (RTPS) and offers Quality of Service(QoS) mechanism. The main differences between the relevant protocols and middleware are summarized in Table 2 [11].

Our middleware services wrap around our customized DDS. It supports publish/subscribe and request/response types of communications. DDS communication has been highly optimized to adopt zero-copy, shared memory approach to transfer data with constant latency, regardless of the size of the payload. Only data of interest is published to a topic and transmitted via middleware communication layer. Each topic can also be assigned a set of QoS descriptors, and ensures reliability, security and meets the storage requirements.

Table 2. Comparison of relevant protocols and middleware for automotive SOA, source [11].

| | | SOME/IP | DDS | eSOC [38] | Signal-oriented |
|---|---|---|---|---|---|
| Communication patterns | Publish-subscribe | x | x | x | |
| | Fire & Forget | x | x | x | |
| | Request-Response / RPC | x | x | x | (x) |
| | Static subscribe | | | | x |
| Discovery / path configuration | Fully static | | | | x |
| | Link to server | x | | | |
| | Link to data | | x | x | |
| Supported protocols | CAN | | | x | x |
| | TCP/IP | x | x | | |
| | UDP/IP | x | x | | x |
| Security | Authentication | | x | | |
| | Encryption | | x | | |
| | Access control | | x | | |
| | Security Tagging | | x | | |
| | Brokerless | | x | x | x |

Our middleware also integrates Client/Server(C/S) and Pub/Sub inside a single binding, and integrates common frameworks including ROS2, and AUTOSAR Adaptive. C/S enables diagnostics, management,and configuration, while the Pub/Sub provides real-time communication. The middleware features "plug-and-play" for applications through dynamic discovery functionalities.

## 7.4 Functional Software

Functional software is the core component of our digital software foundation. It is a data driven, flow-based processing engine. It provides data ingestion, data abstraction, data flow control, and basic services. One of the important tasks of functional software is the functional tasks orchestrations, this includes tasks grouping, binding, static or dynamic configurations, scheduling and executions. Functional software module also monitors and manages the life-cycle of these tasks.

On top of the data flow pipeline, the basic service models and algorithms framework are provided including the environment model, planning model, control model, data management model, functional service model etc. and application developer can utilize the data flow control to acquire sensor data, for example camera data, and use those basic models to design and develop customized applications.

The main data-flow should be capable of handling large amounts of data efficiently and meets the data processing and transport requirements, especially for ADAS and L3/L4autonomous driving applications. The potential performance bottlenecks will be removed including memory copy, data serialization and desensitization, which usually result in optimization of the underlying middleware.

The main data-flow is shown in Figure 10, after data is acquired, data abstraction and data pre-processing steps are performed,and data is then consumed for services, including algorithm services, security service and data management service etc.

The other requirement for functional software is to be able to provide elastic and scalable services,this is necessary to support dynamic nature of applications. Service models should be implemented in a modular, decoupling, configurable and up-gradable fashions.

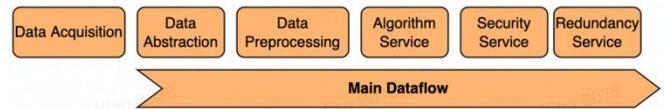

Figure 10. The Main Data-flow.

The function software general framework provides data-flow orchestration, and allows automated configuration, coordination,and management of services. On top of the function software general framework, intelligent driving general framework is created with environment model, planning model and control model, and provides SDKs/APIs to allow OEM/Tier 1s user to develop applications and add new ADAS/AD or connectivity features.

## 7.5 Application Software

The application software layer includes different kinds of ADS or non-ADS applications, such as smart cockpit HMI, ADAS/AD features, vehicle networking connectivity, vehicle cloud collaborations, energy conservation, information,and data security etc. OEMs and other application developers can use the different SDKs provided by the digital foundation platform; they can include but not limited to

a) System or platform SDKs

System or platform SDK can include standardized or formalized data abstractions for hardware components, drivers, supporting software libraries, tool-chains, and componentized OS services. This should cover existing mainline hardware sensors and hardware platforms to be useful and is a continuous process to provide more complete lines of hardware coverage.

b) Environment Model SDKs

Environment model SDK provide all necessary information to support L0-L4 autonomous driving. The sources for environment model SDK can come from the output of perception, localization, fusion and other AI algorithms. It can also comefromV2X and other cloud control sources. Information inside the environment model SDK are classified, normalized and stored in local or remote databases, APIs are provide to enable those similar to Create, Read, Update, and Delete (CRUD) operations for databases, moreover, searches based on fuzzy logic or NLP should be provided to enable users to conduct more powerful and complex searches. For example, user can use "tunnel" "on" "highway" "in" "rain" to pull out all related records in environment model database to construct complex driving Operational Design Domains (ODDs).

c) Configurations SDKs

Configuration SDK contains necessary configurations to develop applications, including parameters for autonomous driving algorithms, service typologies, policies,and other service-related configurations. The schema of configurations is also standardized or formalized.

d) Algorithm SDKs

Algorithm SDK contains information of autonomous driving algorithms, including perceptions, fusion, planning and control. It contains algorithm executable, tool-chains path and information that the loader needs to load and start the algorithms.

e) Mode Management SDK

Mode management SDK acts as launching pad for exiting automated driving system (ADS) functional implementations, often using mature finite state machines (FSMs). It can also act as system mode management to coordinate different FSMs for their state correlations and interactions.

f) Application SDKs

Application SDKs include necessary software libraries and tools that support non-ADS applications, such as data collections, OTA, data security and cloud collaborations.

Using these SDKs, OEMs and applications developers can either develop new applications or customize and enhance their existing applications.

The development of software applications is through SDK, which provides service modeling, data abstraction, and interface normalization, as shown in Figure 11. The SDKs are provided through modular and decoupled service interfaces. Application developers from OEMs and Tier Ones can utilize the SDKs to either enhance the current applications or build their application software of ADAS/AD or cockpit,examples like ACC, LKS. For example, for the ACC (Adaptive Cruise Control) development, application developers can utilize the environment model SDK and system SDK to subscribe to the related sensors, get perception and localization information from environment models, and then feed into the plan and control models, and customize the application development.

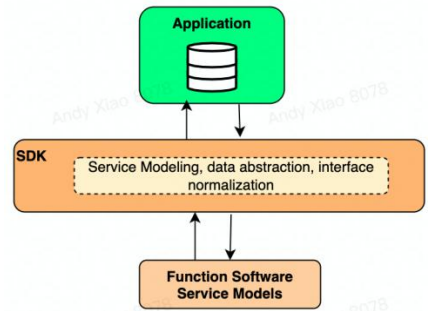

Figure 11. The application interfaces the function software service models through SDK.

## 8. Advantages

From the above, we can see the proposed that the digital foundation platform with multi-layered SOA provide easy development and migration using one system and software architecture, reusing software modules and services for OEMs and other application developers to quickly integrate and adopt different hardware, middleware, and other applications on different platforms.

a) Hardware and software decoupling:

By decoupling hardware and software,each layer can focus on its own functions and services, this is similar to the network TCP/IP stack where each layer (e.g., MAC, IP and TCP) is trying to provide services, address the issues within its own layer without having to worry about the layer underneath or above. This has benefit of rapid development process, upgrading and migration.

b) Scalable and easy to plug-in new hardware:

Through system and platform SDKs, also through HAL layers, support for new hardware(e.g., sensors, actuators, vehicle platforms) can be relatively easily done. Data ingestion interfaces are standardized and formalized, APIs and necessary software libraries and tool-chains are also provided to help new hardware adoptions.

c) Upgradable and reusable software modules:

Through modular and layered design, functional and software modules can be upgraded or reused to migrate and support new hardware, middleware, or services. This helps to reduce the total development cost and improve the Return on Investments (ROIs). Also through service abstractions, it is becoming easier to adapt different flavors of system software and middleware.

Overall,the multi-layered SOA architecture allows software development independent the hardware of the vehicle and improve the agility and speed for OEM developing new vehicle models. The hardware is abstracted, and the digital foundation platform separates the application software from the underneath functional and hardware services. The digital foundation platform also provides POSIX and real time OS, and supports AUTOSAR Adaptive Platform and Classic Platform, and meets requirements of functional redundancy and functional safety. Additionally, the multiple Software Development Toolkits are provided to help OEMs or Tier 1s build application software and define vehicle differentiation.

## 9. Conclusion

The article gives an overview of the automotive intelligence for ACES vehicles, the developments, migration of modern new architectures, AI techniques and methods used for the integration of autonomous functions in next-generation vehicles.

In this paper, we introduced the digital foundation platform, a multi-layered service-oriented architecture which abstracts the hardware and separates the vehicle from the application software at multiple layers. It supports and adapts to vehicle variants, sensors, heterogeneous chip-sets, OS services, middleware services, functional and application developments by providing data, interfaces, and service abstractions between the boundaries of these layers. The goal is to support different sensors, actuators, vehicle platforms at hardware level, to support different operating systems, different middleware using one system and software architecture, decouple hardware and software as well as software development and system integration. This enables fast and easy adoptions and integration at different layers depending on the different needs, reduce the cost for new hardware, software and platform upgrade and integration thus powers OEMs or Tier Ones to continuously update and upgrade the vehicle with application software, data, and AI features.

## 10. Future Work

### Service Granularity

Automotive SOA shall enable the reuse of services across the vehicle generation or ranges (entry, mid, high tier) [25]. The services should be simple and modular enough and the combination of these services to create applications can become large or complex. How fine should the grain services should be carefully considered; high-level applications might not be able to register large number of services. On the other hand, each single service can be, by configurations, policies or ODD requirements, groups, orchestrated together to support more complex autonomous driving functions and applications.

### Data Abstractions

Data abstraction integrates data from different sources, currently the sensor types that the digital foundation platform supports including camera, radar, Lidar, HDMap, GPS, IMU as well as virtual sensors from V2X etc., and there are still more sensor types to support, for example, 4D radar etc. The more complete sets of sensors it supports, the more easier for OEM and other users to quickly bring up and integrate their own hardware with the foundation platform software.

### The Security Firewalls

Firewall acts as a gateway that controls data traffic between different networks. Firewall technologies have been used in traditional ICT for decades, to limit the shortcomings of communication protocols through additional inspections [11]. Risks arise if the persistent protocol weakness is exploited, and the additional controls are bypassed. For automotive, it might be necessary to combine firewalls with secure protocols to ensure information security assets such as confidentiality, integrity, authenticity, and availability on different SOA layers.

## Contact Information

David Yu, VP of Engineering and Chief Architect, AICC Inc.
Email: davidyu@aicc-corp.com

Xinhua Xiao, Ph.D., Software Architect, AICC Inc.
Email: andyxiao@aicc-corp.com

## Abbreviations

| | |
|---|---|
| ACC | Adaptive Cruise Control |
| ACES | Autonomous Driving (Intelligence), Connectivity (Interconnection), Electrification and Shared-Mobility |
| AD | Autonomous Driving |
| ADAS | Advanced Driver Assistance Systems |
| ADS | Automated Driving System |
| AI | Artificial Intelligence |
| AP | Adaptive Platform |
| API | Application Programming Interface |
| ASIC | Application-Specific Integrated Circuit |
| ASIL-D | Automotive Safety Integrity Level D |
| AUTOSAR | AUTomotive Open System ARchitecture |
| CAN | Controller Area Network |
| CP | Classic Platform |
| CRUD | Create, Read, Update, and Delete |
| C/S | Client/Server |
| DCU | Domain Control Unit |
| DDS | Distributed Data Service |
| DFP | Digital Foundation Platform |
| DL | Deep Learning |
| ECU | Electronic Control Unit |
| E/E | Electrical/Electronic |
| FPGA | Field-Programmable Gate Arrays |
| FSM | Finite State Machine |
| HAL | Hardware Abstraction Layer |
| HMI | Human-Machine Interface |
| HPC | High-Performance Computer |
| ICT | Information and Communications Technologies |
| ICV | Intelligent Connected Vehicle |
| LKS | Lane Keeping System |
| LIN | Local Interconnect Network |
| MAC | Media Access Control |
| MCU | MicroController Unit |
| MDC | Mobile Data Center |
| MOST | Media Oriented Systems Transport |
| NLP | Natural language processing |
| NN | Neural Network |
| OEM | Original Equipment Manufacturer |
| OSAL | Operating System abstraction layer |
| POSIX | Portable Operating System Interface for uni-X |
| OTA | Over-the-air |
| QoS | Quality of Service |
| ROI | Return on Investment |
| ROS2 | Robot Operating System 2 |
| RPC | Remote Procedure Call |
| RTOS | Real Time Operating System |
| RTPS | Real Time Publish Subscribe (RTPS) |
| SDV | Software Defined Vehicle |
| SDK | Software Development Toolkit |
| SOA | Service-Oriented Architecture |
| SOME/IP | Scalable service-Oriented MiddlewarE over IP |
| V2V | Vehicle-to-vehicle |
| V2X | Vehicle-to-everything |
| VM | Virtual Machine |